\documentclass[12pt]{article}

\usepackage{graphicx}

\title{  Fractional dynamics of fermion generations}
\author{  Yu.A.Simonov \\
State Research
Center\\Institute of Theoretical and Experimental Physics, \\
Moscow, 117218 Russia}
\newcommand{\beq}{\begin{eqnarray}}
 \newcommand{\eeq}{\end{eqnarray}}
\newcommand{\be}{\begin{equation}}
 \newcommand{\ee}{\end{equation}}

\def\ga{\mathrel{\mathpalette\fun >}}
\def\fun#1#2{\lower3.6pt\vbox{\baselineskip0pt\lineskip.9pt
\ialign{$\mathsurround=0pt#1\hfil ##\hfil$\crcr#2\crcr\sim\crcr}}}

\newcommand{{\SD}}{\rm SD}

\newcommand{{\Mc}}{\mathcal{M}}

\newcommand{\vex}{\mbox{\boldmath${\rm x}$}}

\newcommand{\vep}{\mbox{\boldmath${\rm p}$}}

\newcommand{\vek}{\mbox{\boldmath${\rm k}$}}

\newcommand{\lan}{\langle}
\newcommand{\ran}{\rangle}

\begin{document}
\maketitle
\begin{abstract}
The dynamics of fermion generations is  treated in the  framework  of  the  fractional Lagrangian and Hamiltonian in the   compact extra dimension. The  resulting spectra for $u,d,e$  generations  are described  by the only mass parameter and    slightly different  fractional numbers, while the  neutrino spectra need  another mass parameter  and  a single fractional number for all neutrino species. New  definitions of CKM and PMNS mixing matrices with   standard data are given as the   interference integrals  in the extra dimension. A possibility of higher fermion states as dark matter  candidates is shortly discussed.
 
 \end{abstract}

 \section{ Introduction}

 Fermion masses display a highly hierarchical order, where the mass ratios
 in the up sector can  be larger than  10$^3$ , while in the  down sector of
 the  order of $10^2$, and less in the neutrino sector.

 At the same time all attempts to find   any sign of the internal fermion
 structure have failed and yielded only the lower limit of the internal scale
 of the  order of several TeV \cite{0*}.

 The standard flavor theory \cite{1}  (see also the  recent  reviews \cite{2*,2**,2***}) contains more than 20 parameters, which
 are adjusted to explain the experimental data,    and to suppress  the unobserved
 processes.

 One example is the absence  of the terms yielding the FCNC type reactions (e.g. $\bar \psi_u \hat Z \psi_c)$, in
 the leading approximation.

 It is the main purpose of this paper to address three important  points of the
 flavor theory: 1) what is the basic dynamics of the flavor spectrum, which
 makes it so high hierarchical,  2) why FCNC events are absent in the leading
 approximation, 3) what  is the dynamics, which generates CKM matrix. As will be seen, our approach using the  fractional  dynamics in
 an extra dimension, provides  unexpected results with interesting  possibilities.

 One of those is a possible explanation of the dark matter as the
 higher states of the extra dimension.

 These features, as well as different patterns of fermion mixings in the quark
 and neutrino sectors are subjects of  numerous investigations in the framework
 of  the 4d  Standard Model (SM) (see  \cite{5a,5aa} and references therein).

 Following the idea of the compact extra dimension  \cite{2}, more recently the
 focus of the  analysis turned to theories in extra dimensions
 \cite{3,3*,4,5,5*,6,7,8} e.g. in the five dimensional anti-de Sitter $(Ad S)$
 geometry, as well as in  the modified warped extra dimensions \cite{8,8*,8**}.

 One of the main motivations of these studies was the idea to connect the
 Planck and SM scales in one general approach, using the exponential factors
 differing  the Plank scale from the SM  scales.

 At the same time the presence of the new dimension yields a possibility of  a
 new dynamical mechanism for the  creation of new states -- the Kaluza-Klein
 (KK) states \cite{2}, and  a possible danger, since an infinite tower of KK
 states may appear in the sector under investigation.

 At  the moment there are successful models, generating  fermions and realistic
 mixing at the price of introduction additional parameters \cite{8,8*,8**,10}.

 It is the purpose of the present paper to introduce a simplest possible model,
 exploiting extra dimension with the  first  goal to reproduce the observed
 fermion masses. As will be seen, with the introduction of one mass scale
 parameter $(\mu)$, common to charged leptons, up and down quarks, and one
 dimensionless  fractal  parameter $\gamma$, one is able to reproduce
 fermion masses in all three sectors, slightly varying $\gamma$ in these
 sectors. The neutrino sector requires a different pair  of  value   ( $\mu$, $\gamma$)  for all three species.

To achieve this goal we introduce the path integral formalism in $ 5d$, where
the extra (fifth) dimension is dynamically independent from the 4d, and serves
only to generate fermion masses. This is  similar to the KK mechanism, but
differs in the metric and boundary conditions,which produce a specific
hierarchical spectrum.

In our approach no connection to  gravitation and Planck scale is
contained, but the main emphasis is done on the need of an extra
mechanism for the mass generation, in addition to the  existing
ones in 4 d: Higgs-type mechanism, using vacuum condensate and the
confinement mechanism, using vacuum correlators \cite{9}. The
latter ensures almost 99\% of the mass in the visible part of the
Universe, since it gives mass to all hadrons, including baryons.
The rest of the mass, the 1\% of it, is due to fermions, and this
is  the goal of  the paper to find the source of it, different
from the possible Higgs mechanism, unable at present to predict
masses explicitly. At the same time the invisible (dark) part of
matter is  roughly 6  times larger and we tentatively  envisage for it  the same source
of the extra (fifth) dimension, where excited  generations lack 4d
charges and participate only in  gravitation.

There are several reasons to search for external sources of mass generation.
Indeed, the high masses of  heaviest quarks  would imply a strong  interaction
to be created within 4d. As  an example, the mass generation theory based on
the rainbow version of the  Dyson-Schwinger equation \cite{10},  where all SM
fields participate, would require in all sectors much  stronger couplings than
available in theory. In simple  words, to be as massive as 175 GeV, as the $t$
quark is, and interacting only via QCD and  reasonable (mass-independent) Higgs
couplings seems to be not selfconsistent.

On the other hand, with the extraneous mass generation mechanism,
fermions can be considered  in our 3d space as elementary objects, with no internal
structure, connected with the  internal mass distribution, which
might explain the failure of  all attempts to find that structure
\cite{0*}. Therefore it seems necessary to separate the 4d
dynamics with all SM interacting fields, and the mass generation
process, which may occur via other degrees of freedom, and in this
paper we take the 5-th dimension  to be responsible for  the  mass
generation alone, with no interaction between the 5-th and 4d
coordinates (however with SM interaction in 4d of the
mass,  created in the 5-th coordinate).

At this point one  meets with an unexpected difficulty, of  the
growth of the mass eigenvalue spectrum with the number $n$ in the
ordinary  dynamics as $m_n \sim n^\alpha$ with $\alpha \sim 4 \div
8$. It seems to be impossible with all known types of interaction
in the fifth  dimension (unless one uses different parts of
space-time with  different metrics). Therefore one has to  leave
the familiar formalism of  field theory in $D$  dimensions and
look  elsewhere.

We shall use below the  so-called fractional dynamics, which is
well-developed in many fields, see \cite{15,15*,15**} for books and
reviews. In this formalism the kinetic terms (in our case in the
fifth dimension) acquire a non-integer power form, which creates a
much more versatile character of dynamics. Below in the next
sections we  demonstrate the new form of the fermion spectrum,
generated by the fractional kinetic term, which seems to agree
well with the experimental data.

 The main objective of this  paper is to  suggest a mechanism of  external mass
creation,  which ensures the relativistic fermion spectrum with a minimal
theoretical input. As a  second step we investigate the  mechanism  of flavor
mixing and obtain the  forms of  resulting CKM and PNMS  matrices. In doing this we introduce the new form of those matrices  as the overlap integrals in the 5th dimension of the generation eigenfunctions. Correspondingly in the standard term of the Lagrangian, e.g. $\bar q'_u \hat W q_d$  the  4d integral is extended to 5d. It is shown, that the resulting CKM and PNMS matrices   can be made  realistic.

The paper is organized as follows. In the next section we derive the dynamical
equation for the fermion mass eigenvalues, leaving details of derivation to the
Appendix.  In section 3 we formulate a simple model,  which imitates the same results for the fermion masses, as in the fractional dynamics approach.  In  section 4 the results for the fermion masses are compared with
the data and  the values of few parameters are fixed. In section 5 we discuss
the  problem of the FCNC and show, how it is resolved when  the violating terms are 
defined via the wave functions in the fifth dimension. The concluding section
is devoted  to the  discussion of results.

\section{Fractional dynamics}

The standard dynamics, based of the  quadratic time derivative of
coordinates or the field variable in the  quantum mechanics and
quantum field theory respectively, was  an object of modifications
for different  purposes.

In one case the aim was to extend the dynamics to the cases, when
the system experiences chaotic motion,  diffusion processes etc.,
see e.g. \cite{15,15*, 15**} and  the   references therein.  The
typical for this case is the  resulting kinetic term of the  type
$\left(\frac{\partial z}{\partial t}\right)^\alpha$  or
$(\dot{\varphi})^\alpha$ with $0\leq\alpha \leq 2$.

 At the same time the problem of nonstandard power $\alpha$ of
 time derivatives, and  consequently of the same power $\alpha$ in
 the  field propagator $p^\alpha$, discussed below in the paper,
 was studied in the    framework of the  quantization  of quantum
 gravitation, see e.g. \cite{22} and the  recent review 
 \cite{23}.   In  this case $\alpha$  was   considered to be
 equal to 4, and the main problem was to  circumvent   the
 Ostrogradsky analysis, which in the standard Hamiltonian approach
   yields the  spectrum  instability
 \cite{24}.

 The way out of this problem was recently suggested in
 \cite{25}, where it was shown, that using the path integral
 approach instead of the classical Hamiltonian formalism, one ends
 up with the results, which do not show fatal instabilities, and
 display the reasonable spectrum.

 In what follows we shall use the relativistic path integral
 formalism in 5d, where in the fifth dimension the time derivative
 can have a fractal character and the power $\alpha$ may be
 noninteger and larger than 2.

 We shall be using the Fock-Feynman-Schwinger path integral
 suggested in \cite{9,12,13} For another forms see e.g.
 \cite{18,19}.

 Following this method, the Green's function of a scalar  particle
 $\varphi(x)$ in the $N$  dimensional Euclidean space-time can be
 written as
 \be   G(1,2)  = \left\lan  1 \left|
 \frac{1}{m^2_4-\partial^2_\mu}\right| 1\right\ran = \int^\infty_0
 ds (Dz)_{xy} e^{-K},\label{q1}\ee
 where $K$ is  the kinetic factor playing the role of path
 integral  Lagrangian, and  $(Dz)$ is the path measure
 \be K_N = \int^s_0 d \tau \left(m^2_4 + \sum^N_{\mu=1}
 \frac{1}{4} \left( \frac{\partial
 z_\mu}{\partial\tau}\right)^2\right)\label{q2}\ee
 \be (Dz)_{12} = \int \frac{ dp^{N-1}}{(2\pi)^{N-1}} \prod
 \frac{d^{N-1}\Delta z(n)}{(4\pi \varepsilon)^{N/2}} e^{ip(\sum_n
 \Delta z(n)- (x(1)-x(2)))}\label{q3}\ee

 Here $x(1)=\vex_1; x(2) =\vex_2$.

 In the  case of one fractal  dimension, e.g. $N=4+1$, and
\be m^2-\partial^2_\mu \to m^2 - \sum^4_{\nu=1} \partial^2_\nu -
 (\partial^2_5)^\alpha \mu^{2-2\alpha},\label{q4}\ee
 where $\mu$ is a new scale parameter, the form $K_N$ changes, as
 shown in the Appendix, to
 \be K_4+K_5= \int^s_0 d \tau \left(m^2_4 +  \frac14
 \sum^4_{\mu=1} \left( \frac{\partial z_\mu}{\partial\tau}
 \right)^2 + {\rm const} \left( \frac{dy}{d\tau}\right)^\beta\right),
 ~~\beta=\frac{2\alpha}{2\alpha-1}.\label{q5}\ee

 The basic  transformation, which  is needed to find the spectrum
 in the 5-th dimension,  yields the Hamiltonian form, which in the
 leading Fock amplitude can be written as \cite{12,13}

 \be G(1,2)  = \sqrt{\frac{T_4}{8\pi}} \int^\infty_0
 \frac{d\omega}{\omega^{3/4}} \lan 1| e^{-H(\omega)
 T_4}|2\ran\label{q6}\ee
 where
 \be H(\omega) = \frac{\vep^2 + \mu^{2-2\alpha} p^{2\alpha}_5 +\omega^2+m^2_4}{2 \omega} {2\omega},~ p_5=\frac{\partial}{i\partial
 x_5},x_5 \equiv y.\label{q7}\ee

We assume, that the $y $ dependence is always separated
(factorized) in the resulting wave functions,

\be \psi (\vex, y) =  \chi (\vex) \varphi(y ),\label{8}\ee so that
$\varphi(x_5)$ satisfies equation

\be p^2_5 \varphi (y) =- \frac{\partial^2}{\partial y^2 } \varphi
(y) = m^2_5 \varphi (y ),~~ \mu^{2 - 2\alpha} p_5^{2\alpha}\varphi
= \mu^2 \left(\frac{m_5}{\mu}\right)^{2 \alpha} \varphi
(y).\label{9}\ee

Note, that in (\ref{q6}) one can write following \cite{12,13} for
4d \be \lan \vex | e ^{-H(\omega)T}|\vex'\ran= \sum_k \lan \vex |
k\ran  e^{- E_k (\omega) T} \lan k |\vex'\ran\label{14}\ee where
$k$ runs over discrete and continuous levels of $H(\omega)$. In
 the simplest  case of no 4d interaction one has  \be \lan \vex | k \ran = \chi_k
(\vex) = \frac{1}{(2\pi)^{3/2}} \exp (i\vek\vex).\label{15}\ee

In  the 5d case we add in notations (\ref{14}) the  
\be \lan \bar x | e^{ - H(\omega) T} | \bar x'\ran =
\sum_{k,n} \chi_k (\vex) \psi_n (y) e^{-E_{kn}
(\omega) T} \chi^*_k (\vex') \psi^*_n(y')
  \label{16}\ee
and the  Green's function acquires the form \be  G(1,2)  =
\sqrt{\frac{T_4}{8\pi}} \int^\infty_0
 \frac{d\omega}{\omega^{3/4}}  \int \frac{d^3\vep}{(2\pi)^3} \sum_{n=1,2,3} \psi^+_n (1) e^{-E_n(\omega)
 T_4 +i\vep(\vex_1-\vex_2)} \psi_n(2)\label{q13}\ee
e.g. $\psi_n(1) = \psi_n(x_5)$ is the  eigenfunction of the
Hamiltonian \be H_5=\mu^{2-2\alpha} p_5^{2\alpha}+U(x_5),\label{q15}\ee  where $U(x_5)$ is a possible interaction in $x_5$, which we disregard below.

To go on-shell, one finds the extremum of $H(\omega)$ as a
function of $\omega$, and assuming \be (p^2_5)^\alpha \psi_n =
(m^2_5(n))^\alpha \psi_n\label{q16}\ee one obtains \be E_n =
\sqrt{\vep^2+m^2_4 +\mu^{2-2\alpha}m_5^{2\alpha}
(n)}\label{q17}\ee

In this way each fermion obtains its individual wave function
$\psi_n$, which  accompanies it from the beginning to the  end of
the process without change, unless in the total Lagrangian appear
the terms, which contain integration over $dx_5$, as will be shown
in the next chapter.

At this point we take into account the spinor character of the
fermion and write $S(1,2)$ instead of $G(1,2)$ following the
formalism of \cite{13}.

\be S(x,x') = \left( \frac{1}{m_4+ \hat \partial}\right)_{xx'} = (m_4 - \hat \partial)_x \sqrt{\frac{T}{8\pi}} \int^\infty_0 \frac{d\omega}{\omega^{3/2}} \lan \bar x | e^{-H(\omega) T} | \bar x'\ran.\label{17f}\ee
One can define the covariant  derivative in 5d as \be \hat
\partial = \sum^4_{\nu=1}\partial_\nu \gamma_\mu + \Gamma_5
(\partial_y)^\alpha, ~~ \Gamma_5 = \left(
\frac{1}{i}\right)^\alpha \mu^{1-\alpha}, \label{6'}\ee where
$\mu$ is a mass dimension in the fifth coordinate.
 Here
$\vex, \vex'$ include only 3d spacial coordinates and the
fifth coordinate is $y$ namely, $\bar x = \vex, y,~~ \bar x' =
\vex', y $, and $m_4$ implies the mass generated in 4d (if any).

Finally to define the spectrum due to the 5 dimension,
one  can
write,  assuming periodic boundary conditions

\be \mu^{2-2\alpha} p^{2\alpha}_5 \psi_n (y) = (\pi n)^{2\alpha}
\mu^2 \psi_n (y).\label{q19}\ee which obtain with the wave
functions \be \psi_n (y) = \sqrt{\frac{2}{\pi}} \sin (n\pi y\mu),
~~ \mu y\in [0,1]\label{q20}\ee which yields the spectrum \be m_5 (n)
= \mu(\pi n)^\alpha, ~~ n=1,2,3,...\label{q21}\ee

As will be seen in the next section this type of spectrum can be
useful in comparison with experimental fermion masses.

\section{ An equivalent simple model}

As a simple alternative, one can  consider the interval in the $x_5\equiv y$ space, which depends on the energy (mass), which is generated  on this interval, namely
\be 0\leq y \leq y_+(m); ~~ y_+(m) = \frac{y_0}{ \left( \frac{m}{\mu} \right)^\gamma}. \label{3.1}\ee

Again assuming the mass equation, but now with the standard $p^2$ term, 
 \be p^2 \psi_n = m^2_n \psi_n , ~~ \psi_n = \sqrt{\frac{2}{\pi}} \sin \left( \frac{\pi n y}{y_+ (m_n)}\right),\label{3.2|}\ee
 one arrives at the mass relation
 \be m_n^2 = \frac{\pi n}{y_0} \left(\left( \frac{m_n}{\mu}\right)^\gamma \right)^2\label{3.3}\ee
 and choosing $y_0 = \mu^{-1}$, one has 
 \be m_n = \mu (\pi n )^{\frac{1}{1-\gamma}}, n=1,2,3.\label{3.4}\ee
 
 Comparing this result with (\ref{q21}) in the previous section, obtained within the fractional dynamics approach, one realizes, that 
 \be \alpha = \frac{1}{1-\gamma}.\label{3.5}\ee
 
 This derivation of course does not tell anything about the possible values of $\alpha$ and $\gamma$, however one cannot exclude here the values of  $\gamma>1$, which imitate not growing, but rather decreasing with $n$ spectrum. The latter can be appropriate   e.g. for neutrinos, where one can have $m_{\nu_1} > m_{\nu_2} > m_{\nu_3}>...$
 
 To conclude  this section, it is interesting to check the uncertainty principle $\Delta p \Delta x \ga 1$, which in our case looks like \be\Delta m \cdot \Delta y \to m_n y_+ (m_n)= \left( \frac{m_n}{\mu}\right)^{1-\gamma} = \pi n, \label{3.6}\ee which is strangely  reminiscent of the space quantisation. 

\section{ The 5d fermion spectrum}

From  (\ref{q19})  one has $m_5 (n)= m_n=   \mu( \pi n)^\alpha,
  n=1,2,3,...$ and   writing $\alpha$ in terms of the  power $\gamma$ in (\ref{3.4}), one  has
\be m_n = \mu \cdot (n\pi)^{\frac{1}{1-\gamma}}, ~~
n=1,2,3,...\label{25}\ee

 To
simplify matter, we choose $\mu = 10^{-4}$ MeV, and find for each
fermion the corresponding $\gamma_n$ and $ \alpha_n$  from  (\ref{3.5}), (\ref{25}).

We start with the charged leptons $e,\mu, \tau$ with masses 0.511
MeV, 105.65 MeV and 1776.82 MeV respectively  and taking for them
$n=1,2,3$ in  (\ref{25}), we obtain the values shown in Table 1.

\begin{table}[h]
\caption{ The slope parameters $\alpha_n, \gamma_n$ for leptons $e,\mu,\tau$}

\space{} \begin{center}

\label{tab.01}\begin{tabular}{|c|c|c|c| }\hline
   &$e$&$\mu$&$\tau$\\
\hline
 n& 1&2&3\\
\hline exp. mass(MeV) & 0.511&105.65&1776.82\\

\hline $\alpha_n$&  7.4595&7.547&7.441\\

\hline $\gamma_n$&  0.8659&0.8675&0.8656\\

\hline
\end{tabular}

\end{center}
\end{table}

Note, that if one chooses the universal slope, $(\bar\alpha =7.45410),\bar\gamma=0.8658$ then the 
masses $e, \tau$ are the same within 1\% while the mass of lepton
$\mu$ drops by 18\% $m_\mu \cong 88.65$ MeV.

Here one can expect the danger of appearing the fourth generation charged
lepton for $n=4, m(n=4) =15.5$ GeV. We shall discuss this issue   below in the
last  section, where we assume that this lepton can not acquire 4d charges and hence may play the role of the dark matter particle (as well as higher excited states). 

We turn now to the quark  sector $(d,s,b),$  and keep the same
$\mu =10^{-4}$ MeV. We keep in mind, that the quark masses are scale dependent and usually defined at different scales, e.g. the light quarks at the scale 2 GeV, while $c,b$ and $t$ quark masses are given in the  $\overline{\rm MS}$ scheme. This yields some mass shifts as compared to a unique scale definition, which are smaller than produced by   differencies in $\alpha_n$, $\gamma_n$.  Again from (\ref{25}) we obtain the slope
values $\alpha_n, \gamma_n  , n=1,2,3,$ shown in Table 2.

\begin{table}[!htb]
\caption{ The slope parameters $\alpha_n, \gamma_n$ for quarks $d,s,b$}

\space{} \begin{center}

\label{tab.02}\begin{tabular}{|c|c|c|c| }\hline
   &$d$&$s$&$b$\\
\hline
 n& 1&2&3\\
\hline exp. mass(MeV) & $4.1\div 5.8$&101$^{+29}_{-21}$&4190$^{+180}_{-60}$\\

\hline $\alpha_n$& 9.4520 &7.5226&7.8236\\

\hline $\gamma_n$&  0.894&0.867&0.872\\

\hline
\end{tabular}

\end{center}
\end{table}
 One   can see,  that the $d$ quark and to minor extent  the  $s$ quark decline from
the general slope value. If one chooses $   (\bar\alpha= 7.896450, \bar \gamma=0.8733)$, then  $b$
quark stays within errors, while masses of $d$ and $s$ become
0.842 MeV and 200.6 MeV. At this point one should mention, that
light quark masses are not fixed experimentally with good
accuracy, since are subject to strong interactions, and in
particular the $s$ quark mass inside strange mesons may  be taken 
within an interval of about 100 MeV.

The same procedure for $u,c,t$ quarks, as in previous case leads to the values
shown in Table 3.

\begin{table}[!htb]
\caption{ The slope parameters  for quarks $u,c,t$}

\space{} \begin{center}

\label{tab.02}\begin{tabular}{|c|c|c|c| }\hline
   &$u$&$c$&$t$\\
\hline
 n& 1&2&3\\
\hline exp. mass(MeV) & $1.7\div 3.3$&1270$^{+70}_{-90}$&172000$\pm 900\pm 1300$\\

\hline $\alpha_n$&  8.8465&8.90&9.4796\\

\hline $\gamma_n$&  0.8869&0.8876&0.8945\\

\hline
\end{tabular}

\end{center}
\end{table}

Again, if one fixes $\alpha_n, \gamma_n$ at the $c$ quark value  $(\bar\alpha=8.903, \bar \gamma=0.8876)$,  then the $u$ quark mass stays within experimental limits,
while the $t$ quark mass becomes 3.5 times less, $m_t=47.4 $ GeV.
This means, that effectively the ``trajectory  $\alpha (n)$'' tends
to larger values for larger $n$, and this has an important
consequence for the fate of the fourth (or subsequent) generation.

In an alternative approach one can consider the  $b$ and $t$ quarks in the framework of the strong integration of the third generation with  the Higgs field, which yields the additional mass to both quarks. Indeed, if one keeps the same $\alpha$ for the 3d generation, as for the second, then masses of $b$ and $t$ would be 2.133 GeV and  4.687 Gev respectively and the rest should  be supplied by the Higgs (or else by the new TeV scale physics).

In this case the total mass according to (\ref{q17})~~ is ~~$m_t^2 = m_5^2+m_4^2= \\(172$ GeV$)^2$  with $m_5=46.87$ GeV and $m_4=165.49$ GeV. This latter mass easily fits in the uncertainty intervals of the $H t \bar t$ coupling, found experimentally at LHC \cite{0*}. Therefore with  the unique slope parameter $\alpha_{u,c,t} =8.90$ and $\mu=10^{-4}$ GeV one  obtains  experimental mass values,  $m_u =2.657$ MeV, $ m_c=1.27$ GeV and $m_t= \sqrt{(46.87)^2 + (165.49)^2} = 172$ GeV. In a   similar way keeping the same slope $\alpha_{dsb}=7.5226$ for all $d,s,b$ quarks and the Higgs supported mass of the $b$ quark  $m_4 (b) =3.60$ GeV one obtains $m_s =101$ MeV and $m_b=m_b (\exp)=\sqrt{m^2_4 (b) + m_5^2(b)}= 4.19 $ GeV. Note, that the resulting uncertaintly in the $Hb \bar b$  coupling is again within experimental limits.

Now we come to the neutrino sector and we must first of all  change our
universal scale $\mu$ from $10^{-4}$ MeV to a smaller value. Below we list in
Table 4 the resulting values of slope parameters and neutrino masses.

\begin{table}[!htb]
\caption{ The  neutrino slope parameters  and resulting masses for the
parameter $\mu_\nu$ in (\ref{25}), $\mu_\nu = 3.01\cdot 10^{-6} $ eV.}

\space{} \begin{center}

\label{tab.02}\begin{tabular}{|c|c|c|c| }\hline
   &$\nu_1$&$\nu_2$&$\nu_3$\\
\hline
 n& 1&2&3\\

\hline resulting &&&\\

 masses &&&\\

( $10^{-3}$ eV) &0.434&8.84& 51.5\\

\hline $\alpha_n$&  4.346&4.346&4.346\\

\hline $\gamma_n$&  0.77&0.77&0.77\\

\hline
\end{tabular}

\end{center}
\end{table}

The resulting overall $\bar \alpha_\nu$ is $\bar\alpha_\nu=4.346$, i.e. approximately is twice as small compared to three previous generations.

One can check, that the resulting masses satisfy very  well the   experimental
relations \cite{1,5aa}
$$ |m^2_3 - m^2_2| =(2.44\pm 0.08)  10^{-3} {~\rm  eV}^2$$

$$ m^2_2 - m^2_1 = (7.53 \pm 0.18)\cdot   10^{-5} {~\rm  eV}^2.$$

In this way  the model is able to adjust all fermions with a few parameters.

One can conclude, that the reasonable approximation for the quark
and charged leptons occurs with the only scale parameter
$\mu=10^{-4} $ MeV and one fixed slope parameter for each taste
($e,d,u)$, while only the $t$ quark requires 0.3\% larger  slope
$\gamma_\tau$. Also remarkably the  neutrinos fit in perfectly  well in this
scheme.

\section{ CKM matrix and PMNS matrix  from the  fifth dimension}

In the Standard Model (SM) the flavor Lagrangian  for quarks has
the following  form in the mass basis.
$$ L^{FL} = \bar q_i \hat D q_i + \frac{g}{\sqrt{2}} \bar u^i_L \hat W^+
V^{ij}_{CKM } d^j_L + \bar u^i_L \lambda^{ij}_u u^j_R \left(
\frac{(v+h)}{\sqrt{2}}\right)+$$ \be \bar d^i_L \lambda^{ij}_d
d^j_R \left( \frac{(v+h)}{\sqrt{2}}\right)+ h.c.\label{z18}\ee
where $V^{ij}_{CKM}$ is the CKM matrix.

It is assumed, that the $W$ interaction violates the original
structure of $\hat u, \hat d$ vectors, corresponding  to the $u,d$
mass eigenvalues,  generated by  two last terms in (\ref{z18}),
and this fact leads to the nondiagonal matrix structure of
$V_{CKM}$.

 The last two terms in (\ref{z18})  are basically important for
 the mechanism of fermion masses, since fixation of the  constants
 $\lambda^u_{ij}, \lambda^d_{ij}$ by  the corresponding  values of masses, e.g.
 of $b$ or $c$ quarks allows to predict uniquely   the yield of
 $b\bar b$ and $c\bar c$ quarks in the Higgs decay \cite{2*,2**}.
 However the accuracy of the LHC data in PDG \cite{1} is not enough to
 justify  fully this mechanism. Therefore we
 have suggested  above in this paper another mechanism of mass
 generation -- the  fractional dynamics mechanism and allow the
 Higgs terms  in (\ref{z18}), but consider them as additional  terms
 with undefined  couplings $\lambda^u, \lambda^d$, which can be
 also vanishing, or provide additional mass values e.g. for the third generation ($b,t$).

In what follows we shall  leave this  formalism of SM and instead exploit the
fermion wave function in the  fifth coordinate.

Indeed,  we have 4 sets of eigenfunctions for $\nu, e, d$ and $u$
generations, which depend on the  same fifth coordinate $y$,
defined in  the  fixed region, for example,  $ y_{\min} \leq y\leq
y_{\max}$. These sets we associate with current fermion  states, which may not coincide with mass eigenstates  described in the previous section. 

These four sets can be denoted  as $\chi^{(f)}_i(y), f=e,\nu,d,u. $

Since the fermion masses are given by $m^{(f)}_5(n)$, one  can
neglect in the first approximation any Higgs generated
contribution  and  define the $V_{CKM}^{(5)} $ as \be
V^{CKM}_{ik}  = \int^{y_{\max}}_{ y_{\min}} \chi^{+(f)}_i
(y) \chi^{(f')}_{k} (y) dy.\label{z19}\ee

In the quark case $(f,f') = (u,d)$, and in the lepton case  $(f,f')=(e,\nu)$.

It is important,  that we assign in this way the 5d integral to the matrix
elements  containing   $W$ vertices with fermions.

Instead  all terms with $\gamma, Z$, i.e. the neutral currents
yield the diagonal matrix in $i, k$, i.e.  the FCNC  contribution
is zero in this approximation and  higher order (e.g. box)
diagrams are needed to ensure nonzero results,  as it is supported
by  experimental data \cite{2***,5a}.

One may ask, which properties of the flavor sets produce the observed structure
of the CKM matrix with its almost diagonal form for  $(f, f') =(u,d)$ and the
TBM  form for $ (e\nu)$.

It is clear,  that these flavors differ in their 4d interactions;
not so strongly in the case of $(u,d)$ pair, and  significantly in
the case of ($e\nu)$.  Thus one must deduce, that
$\chi^{(f)}_i(y)$ may be  influenced by the 4d interactions and
flavor symmetries.

One may compare the matrix $V_{CKM}$ with angles between two
orthogonal 3d systems of coordinates,  which are only slightly
rotated with respect to each other in the  case of ($u,d)$ and
strongly rotated in the $(e\nu)$ case.

We start with the CKM mixing matrix, which has the following
structure in the Wolfenstein parametrization \be V_{CKM}^{ij} =
\left( \begin{array}{lll} V_{ud} &V_{us}&V_{ub}\\ V_{cd}
&V_{cs}&V_{cb}\\V_{td} &V_{ts}&V_{tb}\end{array}\right) = \left(
\begin{array}{lll} 1-\frac{\lambda^2}{2}& \lambda&A\lambda^3
(\rho-i\eta)\\-\lambda&1-\frac{\lambda^2}{2}&  A\lambda^2\\
 A\lambda^3
(1-\rho-i\eta),&-A\lambda^2&1\end{array}\right) \label{q24}\ee with the estimates
 \cite{1}
 $$ |V_{ud}| \simeq |V_{cs}|\simeq |V_{tb}|\simeq 1, ~~
|V_{us}| \simeq |V_{cd}|\cong 0.22$$ \be |V_{cb}| \simeq |V_{ts}|\simeq 4\cdot
10^{-2}, ~~ |V_{ub}| \simeq |V_{td}|\simeq 5\cdot 10^{-3}\label{q25}\ee
 and we
are choosing   the set of orthonormal functions  $\chi^{(d)}_i, \chi^{(u)}_k$ denoting $\chi^{(d)}_i (i=d,s,b), \chi^{(u)}_k (k=u,c,t)$  $\chi^{(d)}_i = U^{(d)}_{in} \psi_n (\varphi); \chi^{(u)}_k = U_{kn'}^{(u)} \psi_{n'}(\varphi)$, one has 
\be V_{CKM}^{ij} = \int^\pi_0 \chi^{*(u)}_i \chi_j^{(d)}  d\varphi = (U^{(u)})^*_{in} U^{(d)}_{jn} = U^{(d)} U^{(u)^+})_{ji}.\label{26*}\ee
Assuming, that $U^{(d)}, U^{(u)}$ are unitary matrices, one ensures the unitarity of  the  resulting CKM matrix $V_{CKM}$ e.g. in the simplest example one takes   
\be \psi_n
(\varphi) =\sqrt{\frac{2}{\pi}} \sin  n\varphi, ~~ n=1,2,3,... 0\leq
\varphi\leq\pi, \varphi=\pi\mu y.\label{q26}\ee To produce the mixing one  can use two different
orthonormal sets, which can be constructed from (\ref{q26}). To  simplify the
matter we choose for the $(u,c,t)$ the original set (\ref{q26}) with $n=1,2,3,$
while for ($d,s,b)$ one can assign as a first approximation the set
$$ \chi_1^{(d)} =\sqrt{\frac{2}{\pi}}(V_{11} \sin \varphi + V_{12} \sin 2\varphi +
V_{13} \sin 3\varphi)$$

$$ \chi_2^{(d)} =\sqrt{\frac{2}{\pi}}(V_{21} \sin \varphi + V_{22} \sin 2\varphi +
V_{23} \sin 3\varphi)$$ \be  \chi_3^{(d)} =\sqrt{\frac{2}{\pi}}(V_{31} \sin \varphi +
V_{32} \sin 2\varphi + V_{33} \sin 3\varphi),\label{q27}\ee where $V_{ij}$ are
the same as in (\ref{q24}).

Here the,   orthonormality  of the  set $\chi_i^{(f)}$ is restored, when explicit unitary values of $V_{ik}$ are used in (\ref{q27}), so that the unitarity of $V_{CKM}$ is  supported, when two orthonormal sets of $\chi_{u,c,t}$ and $\chi_{d,s,b} $ are used for the definition of $V_{CKM}$ as in (\ref{q4}).

It is clear, that the resulting CKM matrix given in(\ref{q24}) will be exactly
reproduced by the integrals \be V_{ij} =\int^\pi_0 \chi_i^{(u)^+} (\varphi)
\chi_j^{(d)} (\varphi) d\varphi.\label{q28}\ee

 It is clear, that the
sets of $\chi_i (\varphi) $ can be chosen in many different ways,  should depend on  4d SM quantum numbers.

We now turn to the PMNS matrix $U_{PMNS}$, which we also define in the fifth dimension,
\be \chi_i^{(\nu)} = U^{(\nu)}_{in} \psi_n (\varphi);  ~~\chi_k^{(l)} = U_{kn}^{(l)} \psi_n (\varphi)\label{30}\ee
with the result 
\be U_{ik}^{PMNS}  = \int^\pi_0 d\varphi \chi_i^{(l)+}(\varphi) \chi_k^{(\nu)} (\varphi) = U_{nk}^{(l)+} U_{in}^{(\nu)} =(U^{(\nu)} U^{(l)+})_{ik}\label{31}\ee

 Again, as  in the case of $U^{CKM}$, the unitarity of $U^{PMNS}$ is ensured by the unitary matrices $U^{(\nu}), U^{(l)}$.
 
  We shall be using the standard parametrisation of $U^{PMNS}$ with the  convention of the smallest mixing angle in $s_{13} =\sin \theta_{13}$, \cite{1,5aa}, also assuming the Dirac neutrino nature.
  
  \be U^{PMNS}= \left(\begin{array}{lll}
  c_{12}c_{13}& s_{12}c_{13},&s_{13}e^{-i\delta}\\
  -s_{12}c_{23}-c_{12}s_{23}s_{13} e^{i\delta},& c_{12}c_{23} -s_{12}s_{23}s_{13} e^{i\delta}, & s_{23}c_{13}\\
  s_{12}s_{23}-c_{12}c_{23}s_{13} e^{i\delta},& -c_{12}s_{23} -c_{12}s_{23}-s_{12} c_{23} s_{13} e^{i\delta}, & c_{23}c_{13}\end{array} \right)\label{32}\ee
  with the best fit values \cite{1,5aa}
  $$ \sin^2 \theta_{12} =0.308 \pm 0.017; ~~ \sin^2 \theta_{23} (\Delta m^2 >0) =0.437^{+0.033}_{-0.023}$$\be\sin^2 \theta_{13} (\Delta m^2 >0) =0.0234^{+0.0020}_{-0.0019},~~ \delta/\pi = 1.39^{+0.38}_{-0.27}.\label{33}\ee
  In what follows we choose, as in the CKM matrix case, the more massive lepton matrix $U^{(l)}$ to be diagonal, $U_{ij}^{(l)} = \delta_{ij}$. As a consequence one has $U^{PMNS} = U^{(v)} $. One can see, that even with the simplistic choice of the matrices $U^{(\nu)}, U^{(l)}$ and the orthonormal set of functions $\{ \sqrt{\frac{2}{\pi}} \sin (n\varphi)\}$ allows to reproduce $U^{PMNs}$. In  this way the choice of the fifth dimension as the source of the generation dynamics, and at the same time, of the dynamics behind the flavor mixing, might be of interest for the future development. For instance, the matrices $U^{(\nu)}$ and $U^{(l)}$ depend on the symmetry and interaction in the $\{\nu\}$ and $\{l\}$  generations, or, rather, on the ``projections'' of these symmetries in the $x_5$ space.

\section{Conclusion and discussion}

We have studied above the possible way to obtain flavor hierachical masses from
one scale $\mu$, and the high power kinetic term $p^\alpha$ in the  extra
dimension. Surprisingly the resulting masses for charged leptons and quarks are
described  well with one scale $\mu$ and slightly different slope parameters,
while neutrinos need another scale $\mu$ and unique slope.

We have extended this analysis to calculate the CKM matrix via the overlap
integrals of extra dimensional fermion wave functions and obtain reasonable
results when the full current  sets of $(u,c,t)$ and $(d,s,b)$ as functions of $x_5$   are slightly  different.

It is important, that  we assign in this way  the 5d integral to the matrix elements  containing $\gamma, Z$, and $W$ vetrices with fermions.

As a result all  terms with $\gamma , Z$, i.e. the neutral currents  yield the diagonal matrix in $n,n'$ i.e. the FCNC  contribution is zero in this approximation and higher order (e.g. box) diagrams are needed to ensure nonzero results, as it is supported by  experimental data (see e.g. \cite{2***} and \cite{5a}.

In this picture one obtains, however, many excited fermionic states with  $n>3$,
and  one can associate these states with dark matter, if those do not acquire
nonzero charges in weak, strong and electromanetic interactions, as may be
dictated by symmetry requirements. Indeed one may argue, that three lowest generations are subject to some kind of the  three-fold (replica) symmetry, as it was  assumed before in \cite{34,35,36}, also using extra dimmensions in \cite{37,38,39,40,41}, and recently in the discrete flavor symmetry approach \cite{5aa, 42,43} or else in the three-site gauge model of flavor hierarchy \cite{44}, where the Pati-Salam symmetry model is  considered as  a product $(PS)^3$. In this framework one can treat the higher generations as not connected to the standard  SM symmetries and interactions. The fermions of these higher generations would not have any charges, except for the mass -- the  gravitational charge, and hence behave as particles of the dark matter.  The resulting high mass of these states
($ m^*_u\geq 2.60 $ TeV, $m^*_d \geq 47.8 $ GeV, $m_e^*\geq 15.61$ GeV, $m_\nu^* \geq 0.18 $ eV) 
might explain the mass  dominance of dark matter over the  visible one. The
fundamental ground of the  presented approach,  which would allow to calculate  $ \alpha$ for different  fermions, is still missing and waits for
additional study.

This  work was done in the framework of the scientific program of the Russian
Science Foundation RSF, project number 16-12-10414.

\setcounter{equation}{0} \def\theequation{A1.\arabic{equation}}

\vspace{2cm}
 \setcounter{equation}{0}
\renewcommand{\theequation}{A.\arabic{equation}}

\hfill {\it  Appendix  1}

\centerline{\it\large From  fractional Lagrangian to fractional
path-integral Hamiltonian}

 \vspace{1cm}

\setcounter{equation}{0} \def\theequation{A1.\arabic{equation}}

I.  We start with  a quantum mechanical  example and consider  the 1d Lagrangian

\be  L = \eta(\dot{x}^2)^{\beta/2}- V;~~ p=\frac{\partial L}{\partial \dot{x}}
= \eta \beta \dot{x} (\dot{x}^2) ^{\frac{\beta}{2}-1}\label{A1}\ee and the
resulting Hamiltonian assumes the form \be H= p\dot{x} -L = \eta (\beta -1)
(\dot {x}^2)^{\beta/2} +V = \eta (\beta-1) \left(
\frac{p^2}{\eta^2\beta^2}\right)^{ \frac{\beta}{2(\beta-1)}}+V.\label{A2}\ee
Writing (\ref{A2})  in the form $H=\mu^{1-\alpha} p^\alpha$, one can see that
$\alpha = \frac{\beta}{\beta-1}$ may be however large, when $\beta$ tends to 1,
implying a strong hierarchy of eigenvalues.

II. We now turn to the field theory and define the structure of the kinetic
part of Lagrangian for the scalar field $\varphi (x,y\equiv x_5)$. \be L \to
\left( \frac{\partial \varphi^+}{\partial x_\mu}\right) \left( \frac{\partial
\varphi}{\partial x_\mu}\right)+ \mu^{2-2 \alpha}L \varphi^+ (\partial^2_5)
^{\alpha-1/2} \varphi + (\partial^2_5)^{\alpha-1/2} \varphi^+
\partial_5 \varphi].\label{A3}\ee

This gives for the field propagator \be D_\varphi = \frac{1}{m^2_q
-\partial^2_\mu - \mu^{2-2 \alpha}  \partial_5^{2\alpha}}.\label{A4}\ee

In the same way one obtains the propagator for the  fermion, where the form
(\ref{A4}) appears in the  denominator.

III. We now turn to the path integral form of the  propagator $D_\varphi$,
doing the same transformations as in \cite{13}, but now with
$\partial^{2\alpha}$ instead of  $\partial^2$ in the  denominator. One has using \cite{15**} 
$$ D_\varphi= \lan 1| \int^\infty_0 ds e^{-s(m^2_4 - \partial^2_\mu -
\mu^{2-2_\alpha}\partial_5^{2_\alpha})}|2>=$$
 \be  = \lan 1| \int^\infty_0 ds  (D^5z)_{12} e^{-K}\Phi_{4d} (x,y)
 |2>.\label{A5}\ee

Here $ K=K_4+K_5$ with \be K_4 = \int^s_0 d\tau \left(m^2_4 + \frac{1}{4}
\left(\frac{ d z_\mu}{d t} \right)^2 \right).\label{A6}\ee

To arrive at $K_5$   one should use the procedure, described in \cite{13} for
the standard path integral, see also \cite{15} in the  case of the fractional
derivatives. To this end  consider one step of the path integral in the $x_5$
coordinate \be <y_{k+1} | e^{-\Delta \tau (-\partial^2_5)^\alpha\mu^{2-2
\alpha} }|y_k> = \int dq e^{iq (y_{k+1} -y_k)-\Delta \tau q^{2\alpha}\mu^{2-2
\alpha} }.\label{A7}\ee

Finding the extremum of the integral in $dq$ one  obtains the exponent \be \exp
\left(- \int^s_0 d\tau \mu^2  \left[ -\frac{1}{4\alpha^2} \left( \frac{ dy}{
d\tau } \right)^2\right]^{\frac{ \alpha}{2 \alpha-1}}\right)= \exp
(-K_5).\label{aA7}\ee

One can see, that $K_5$ tends to the form $K_4$ in the limit $\alpha\to 1 $  (
with $m_4\equiv 0$).

To obtain the path integral Hamiltonian one can use the same relations as in
the part I of  the Appendix with $\frac{\alpha}{2 \alpha-1}= \frac{\beta}{2}$
yielding finally \be H_5=A  p_5^{2\alpha}\label{A8}\ee where $A$ is made  of
constants $\alpha$ and $\mu$. Writing finally as in \cite{13} $d\tau =
\frac{dt_4}{2\omega} $one  arrives at the form given in Eq. (\ref{q3}). A more
accurate form, is obtained in the same way as in section  (3.1) of \cite{15},
in the form of the Fox H-function.

\end{document}